\documentclass[aps,pra,showpacs,amssymb,superscriptaddress,nofootinbib,twocolumn]{revtex4}

\usepackage{graphicx}

\begin{document}

\def\half{\textstyle\frac{1}{2}}

\title{Generation of bipartite spin entanglement via spin-independent scattering}

\author{Lucas Lamata} \email{lamata@imaff.cfmac.csic.es}

\affiliation{Instituto de Matem\'aticas y F\'{\i}sica Fundamental,
CSIC, Serrano 113-bis, 28006 Madrid, Spain}

\author{Juan Le\'on} \email{leon@imaff.cfmac.csic.es}

\affiliation{Instituto de Matem\'aticas y F\'{\i}sica Fundamental,
CSIC, Serrano 113-bis, 28006 Madrid, Spain}

\begin{abstract}
We consider the  bipartite spin entanglement between two identical
 fermions generated in spin-independent scattering. We show how the spatial
 degrees of freedom  act as ancillas for the creation of  entanglement to a degree that
 depends on the scattering angle, $\theta$. The number of Slater determinants generated
   in the process is greater than 1, corresponding to genuine quantum correlations between
  the identical fermions. The maximal entanglement attainable of 1 e-bit is reached at $\theta=\pi/2$.
   We also analyze a simple $\theta$ dependent Bell's inequality, which is violated for
    $\pi/4<\theta\leq\pi/2$. This phenomenon is unrelated to the symmetrization postulate
    but does not appear for unequal particles.
\end{abstract}

\pacs{03.67.Mn,03.65.Nk,03.65.Ud}

\maketitle
\section{Introduction}
Bipartite and multipartite entanglement is the main resource that
gives rise to many of the applications of quantum information and
computation, like for example quantum teleportation
\cite{tele1,tele2} and quantum cryptography \cite{crypto1,Eke91},
among others (see for instance Refs. \cite{nielsen,GM02}). A
compound system is entangled when it is impossible to attribute a
complete set of properties to any of its parts. In this case, and
for pure states, it is impossible to factor the state in a product
of independent factors belonging to its parts. In this paper we will
consider bipartite systems composed of two $s=\half$ fermions. Our
aim is to uncover some specific features that apply when both
particles are identical. They appear itemized in the next page.

States of two identical fermions have to obey the symmetrization
postulate. This implies that they decompose into linear combinations
of Slater determinants (SLs) of individual states. Naively, as these
SLs cannot
 be factorized further, indistinguishability seems to imply entanglement. This is reinforced by the observation
  that the entropy of entanglement (EoE) is bounded from below
  by $S\geq 1$, well above the lower limit $S=0$ for a pair of non-entangled distinguishable particles.
  So, it looks like there is an inescapable amount of uncertainty, and hence of entanglement, in any state of two identical fermions.
The above issue has been extensively examined in the literature
\cite{entanglefermion1,ESB+02,entanglefermion2} with the following
result: Part of the uncertainty (giving $S=1$) corresponds to the
impossibility to individuate which one is the first or the second
particle of the system. This explains why the lower limit for the
EoE is 1. Consider for instance two identical $s=\half$ fermions in
a singlet state
$$\chi_S:=\frac{1}{\sqrt{2}}[\chi(1)^{\uparrow}\chi(2)^{\downarrow}-
\chi(1)^{\downarrow}\chi(2)^{\uparrow}].$$ The antisymmetrization
does not preclude the assignment of properties to the particles, but
only assigning them precisely to particle 1 or particle 2. The
reduced density matrix of any of the particles is $\rho =
\frac{1}{2}\openone$ with an EoE $S(\rho) = 1$. The portion of $S$
above 1 (if any) is genuine entanglement as it corresponds to the
impossibility of attributing precise
 properties to the particles of the system \cite{entanglefermion2}. Assume for instance that we endow the previous fermions with the capability of being
  outside ($\chi=\psi$) or inside ($\chi=\varphi$) the laboratory ($(\psi^i,\psi^j)=\delta^{ij}$, $(\varphi^i,\varphi^j)=\delta^{ij}$,
$(\psi^i,\varphi^j)=0$, $i,j=\uparrow, \downarrow$). We now have two
different possibilities: either the fermion outside has spin up
($\psi^{\uparrow}$) or spin down ($\psi^{\downarrow}$). Hence, there
are two different SLs for a system built by a pair of particles
 with opposite spins, one outside, the other inside the laboratory
\begin{eqnarray}
{\rm SL}(1,2)_1 &=&
\frac{1}{\sqrt{2}}[\psi(1)^{\uparrow}\,\varphi(2)^{\downarrow}-
\varphi(1)^{\downarrow}\, \psi(2)^{\uparrow}], \nonumber \\
{\rm SL}(1,2)_2 &=&
\frac{1}{\sqrt{2}}[\psi(1)^{\downarrow}\,\varphi(2)^{\uparrow}-
\varphi(1)^{\uparrow}\, \psi(2)^{\downarrow}]. \label{j1}
\end{eqnarray}
They form two different biorthogonal states, the combination $[{\rm
SL}(1,2)_1-{\rm SL}(1,2)_2]/\sqrt{2}$ corresponding to the singlet
  and $[{\rm SL}(1,2)_1+{\rm SL}(1,2)_2]/\sqrt{2}$ to the triplet state (with respect to the total spin $\mathbf{s}=\mathbf{s}_1+\mathbf{s}_2$).
  An arbitrary state $\Phi(1,2)$ would then be a linear combination of these two SLs:
\begin{equation}
\Phi(1,2)\,=\, c_1 {\rm SL}(1,2)_1\,+\,c_2 {\rm SL}(1,2)_2,\,\,
\sum_i|c_i|^2 = 1, \label{j2}
\end{equation}
giving an EoE
\begin{equation}
S\,=\,1-\sum_i |c_i|^2\, \log_2|c_i|^2 \,\geq \,1. \label{j3}
\end{equation}
Clearly, when $c_1$ or $c_2$ vanish, we come back to $S=1$, as the
only uncertainty left is the very identity of the particles.
 Summarizing, while indistinguishability is an issue to be solved by antisymmetrization within each SL, entanglement is an
  issue pertaining to the superposition of different SLs \cite{entanglefermion1,ESB+02,entanglefermion2}. At the end, we could even decide to call 1 to the variables of the outside
   particle, and forget about symmetrization
\begin{equation}
{\rm SL}(1,2) \rightarrow \, c_1
\psi(1)^{\uparrow}\,\varphi(2)^{\downarrow}\,+\,c_2
\psi(1)^{\downarrow}\,\varphi(2)^{\uparrow}, \label{j4}
\end{equation}
as both particles are far away from each other. In this case, the
EoE $S=\,-\sum_i |c_i|^2\, \log_2|c_i|^2\geq 0$ is lesser than the
one corresponding to antisymmetrized states by a quantity of 1,
which is just the uncertainty associated to antisymmetrization. From
now on we will consider the latter definition of $S$, which gives
the genuine amount of entanglement between the two particles. Notice
that for half-odd $s$, the number $\#{\rm {\rm SL}}$ of Slater
determinants is bounded by $\#{\rm {\rm SL}}\leq (2s+1)d/2$, where
$d$ is the dimension of each Hilbert space of the configuration or
momentum degrees of freedom for each of the two fermions.

Much in the same way as above, we could consider one of the
particles as right moving ($\chi=\psi_0$) the other as left moving
($\chi=\psi_\pi$), giving rise to two SLs in parallel with the above
discussion. This is the first step towards the inclusion
 of the full set of commuting operators for the system. In addition to the spin components ($s_1, s_2$) or helicities, there
  are the total $\mathbf{P}$ and relative $\mathbf{p}$ momenta. In the center of mass (CoM) frame we could consider the system
   described by the continuum of SLs
\begin{eqnarray}
{\rm SL}(1,2;\mathbf{p})_s &=&
\frac{1}{\sqrt{2}}[\psi(1)_0^{s}\,\psi(2)_\pi^{-s}-
\psi(1)_\pi^{-s}\, \psi(2)_0^{s}], \nonumber\\
{\rm SL}(1,2;\mathbf{p})_{-s} &=&
\frac{1}{\sqrt{2}}[\psi(1)_0^{-s}\,\psi(2)_\pi^{s}-
\psi(1)_\pi^{s}\, \psi(2)_0^{-s}], \label{j5}
\end{eqnarray}
where $\psi(1)_0^s=\langle 1|\mathbf{p}\,s\rangle$ and $\psi(1)_\pi^s=\langle 1|-\mathbf{p}\,s\rangle$. The labels 0 and $\pi$
 are the azimuthal angles when we laid the axes along $\mathbf{p}$. Finally, there is a pair of SLs for each $\mathbf{p}$,
 so that a general state made with two opposite spin particles with relative momentum $\mathbf{p}$ could be written in the form:
\begin{equation}
\Phi(1,2)_\mathbf{p}^0\,=\, \sum_{s=\pm 1/2}\,c_s(\mathbf{p})\,{\rm
SL}(1,2;\mathbf{p})_s, \label{j6}
\end{equation}
with $\sum_{s=\pm 1/2}\,|c_s(\mathbf{p})|^2 =1$. Again, we run into
the impossibility to tell which is 1 and which is 2. In addition
there may be some uncertainty about the total spin state, whether a
singlet or a triplet, or conversely,
 about the spin component of  any of the particles, $\psi_0$ or $\psi_\pi$.

After this discussion it should be clear to what extent entanglement
and distinguishability belong to different realms
\cite{entanglefermion1,ESB+02,entanglefermion2}. The only
requirement to include identical particles is to symmetrize the
expressions used for unlike particles. Until now, we have only
considered the free case. We have to examine the case of two
interacting particles, as interaction is expected to be the source
of subsequent entanglement
\cite{LM76,Tor85,PS03,MY04,AAM04,SALW04,H05,TK05,LLS05,W05}.
Obviously, the answer may depend on a tricky way on the detailed
form of the interaction, of its spin dependence in particular . It
also seems that the role of particles identity, if any, will be
played through symmetrization.

In the following we will show that spin entanglement is generated
for the case of two interacting spin-$\half$ identical particles,
with the following features:
\begin{itemize}
\item Spin-spin entanglement is generated even by spin independent interactions.
\item In this case, it is independent of any symmetrization procedure.
\item This phenomenon does not appear for unlike particles.
\end{itemize}
\section{Spin entanglement via spin-independent scattering}
We first tackle the scattering of two unequal $s=\half$ particles
$A$ and $B$ which run into each other with relative CoM momentum
$\mathbf{p}$. We set the frame axes  by  the initial momentum
$\mathbf{p}$ of particle $A$, and let the spin components be $s_a=s$
and $s_b=-s$ along an arbitrary but fixed axis. We will consider a
spin independent Hamiltonian $H$, so the evolution conserves
$\mathbf{s}_a$ and $\mathbf{s}_b$. We denote by $A_{\theta}^s$
$(B_{\theta}^s)$ the state of particle $A$ ($B$) that propagates
along direction $\theta$ with spin $s$. In these conditions the
scattering proceeds as:
\begin{equation}
\Phi_{\mbox{in}}=\, A_0^s\, B_\pi^{-s}\longrightarrow\,
\Phi_{\mbox{out}}(\theta)=\,f_p(\theta) \,A_\theta^s \,
B_{\pi-\theta}^{-s},\label{j7}
\end{equation}
where $\theta$ is the scattering angle and $f_p(\theta)$ the
scattering amplitude. We will consider $\theta$ different from 0 or
$\pi$ to avoid forward and backward directions. While the increase
of uncertainty due to the interaction is clear, because a continuous
manifold of final directions with probabilities $|f_p(\theta)|^2$
opened up from just one initial direction, spin remains untouched.
The information about $s_a$ is the same before and after the
scattering; as
 much as we knew the initial spin of $A$, we know its final spin whatever the final direction is. In other words, spin
  was not entangled by the interaction. We will now translate these well known facts to the case of identical particles,
   where they do not hold true.

\begin{figure}
\includegraphics[width=8cm]{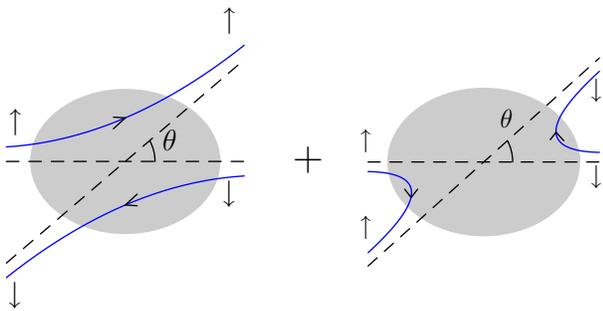}
\caption{(Color online) Schematic picture of the two channels that
contribute to the spin-independent scattering of two identical
fermions. The shaded regions denote an arbitrary spin-independent
interaction between the two fermions. The vertical arrows
$\uparrow$, $\downarrow$ indicate the corresponding third component
of spin.}\label{figsse1bis}
\end{figure}

Let particle $B$ be identical to $A$. Consider the same initial
state as before: A particle $A$ with momentum $\mathbf{p}$
 and spin $s$ runs into  another $A$ with momentum $-\mathbf{p}$ and spin $-s$. Notice there is maximal information on the
  state. We could write $\Phi_{\mbox{in}}=\, A_0^s A_\pi^{-s}$, and eventually symmetrize. We now focus on the final
   state. It is no longer true that particle $A$ will come out with momentum $\mathbf{p'}$ and spin $s$ with amplitude
    $f_p(\theta)$  while the amplitude for coming out with momentum $\mathbf{p'}$ and spin $-s$ vanishes. Recalling
     that $B$ above did become $A$, the two cases
     $f_p(\theta) A_\theta^s B_{\pi-\theta}^{-s}$ and $f_p(\pi-\theta)A_{\pi-\theta}^{s} B_\theta^{-s}$
      fuse into a unique state
\begin{equation}
\Phi_{\mbox{out}}(\theta)=f_p(\theta) \,A_\theta^s\,
A_{\pi-\theta}^{-s}+f_p(\pi-\theta)\,A_{\pi-\theta}^{s} \,
A_\theta^{-s},\label{j8}
\end{equation}
as shown in Figure \ref{figsse1bis}.
Notice the uncertainty acquired by the spin: Now particle $A$ comes
out from the interaction along $\theta$ either with spin $s$ or with
spin $-s$, with relative amplitudes $f_p(\theta)$ and
$f_p(\pi-\theta)$ respectively. In other words, spin was entangled
during the spin independent evolution. Here, it is not the spin
dependence of the interaction, but the existence of additional
degrees of freedom which generate spin-spin entanglement. These, act
as ancillas creating an effective spin-spin interaction that
entangles the two fermions. The ancilla and the degree of
entanglement depend on the scattering angle $\theta$. Notice that
for $\theta=\pi/2$ both amplitudes $f_p(\theta)$ and
$f_p(\pi-\theta)$ become equal, so that the degree of generated
entanglement is maximal, 1 e-bit. On the other hand, for
$\theta\simeq 0$, it generally holds $f_p(\theta)\gg
f_p(\pi-\theta)$, so that in the forward and backward scattering
almost no entanglement would be generated. However, this depends on
the specific interaction. In Sec. \ref{asecslo} we will clarify this
point with Coulomb interaction.

Symmetrization does not change this, it only expresses that we can
not tell which one is  1 and which one is 2. The properly
symmetrized initial state is
\begin{eqnarray}
\Phi_{\mbox{in}}&=&{\rm SL}(1,2;\mathbf{p})_s\nonumber\\
&=&\frac{1}{\sqrt{2}}[A(1)_0^s A(2)_\pi^{-s}-A(1)_\pi^{-s}A(2)_0^s].
\label{j9}
\end{eqnarray}
The scattering process could be written in terms of SLs as
\begin{eqnarray}
&{\rm SL}(1,2;\mathbf{p})_s\longrightarrow &\nonumber\\
&f_p(\theta)\, {\rm SL}(1,2;\mathbf{p}')_s - f_p(\pi-\theta)\, {\rm
SL}(1,2;\mathbf{p}')_{-s},& \label{j10}
\end{eqnarray}
where $\mathbf{p}'$ is the final momentum and the Slater
determinants are given in (\ref{j5}). Both, this expression  and
Eq.(\ref{j8}), describe the same physical situation and  lead to the
same entanglement generation.

The bosonic case may be analyzed in an analogous way. The
modification for two-dimensional spin Hilbert spaces (i.e. photons)
would be a sign change in Eqs. (\ref{j5}), (\ref{j9}) and
(\ref{j10}), as bosonic statistics has associated symmetric states.
The equivalent of Eq. (\ref{j10}) for bosons is a genuine entangled
state for $\theta\neq0,\pi$, much as in the fermionic case.
\section{A specific example: Coulomb scattering at lowest order\label{asecslo}}
We now consider Coulomb interaction at lowest order to illustrate
the reasonings presented above. In this case
\begin{eqnarray}
f_p(\theta) & = & \frac{N(e)}{t(\theta)},\nonumber\\
f_p(\pi-\theta) & = & \frac{N(e)}{u(\theta)},\label{eqsse11}
\end{eqnarray}
where $N(e)$ is a numerical factor depending on the charge $e$.
$t(\theta)$ and $u(\theta)$ are two of the Mandelstam variables,
associated to $t$ and $u$ channels respectively, and depending on
the scattering angle $\theta$. for initial $p$ and final $p'$
relative 4-momenta of the scattering fermions, they are given by
$t=(p-p')^2$, $u=(p+p')^2$. In the CoM frame,
\begin{eqnarray}
 t(\theta)\!\! & := & 2(m^2-E^2)(1-\cos\theta),\nonumber\\
 u(\theta)\!\! & := & 2(m^2-E^2)(1+\cos\theta),\label{eqsse12}
\end{eqnarray}
where $m$ is the mass of each fermion and $2E$ is the available
energy.

According to this, the spin part of the state (\ref{j8}) for this
case, properly normalized, is
\begin{eqnarray}
|\chi_{\theta}\rangle=f_+(\theta)|\!\!\uparrow\downarrow\rangle-
f_-(\theta)|\!\!\downarrow\uparrow\rangle,\label{eqsse13}
\end{eqnarray}
being
\begin{eqnarray}
f_{\pm}(\theta):=\frac{1\pm\cos\theta}{\sqrt{2(1+\cos^2\theta)}}.\label{eqsse14}
\end{eqnarray}
The two amplitudes $f_+$ and $f_-$ vary monotonously as $\theta$
grows, becoming equal for $\theta=\pi/2$. The physical meaning for
this is that for $\theta\rightarrow0$, the knowledge about the
system is maximal and the entanglement minimal (zero), and for
increasing $\theta$ the knowledge of the system decreases
continuously until reaching its minimum value at $\theta=\pi/2$.
Accordingly, the entanglement grows with $\theta$ until reaching its
maximum value for $\theta=\pi/2$.

\begin{figure}
\includegraphics{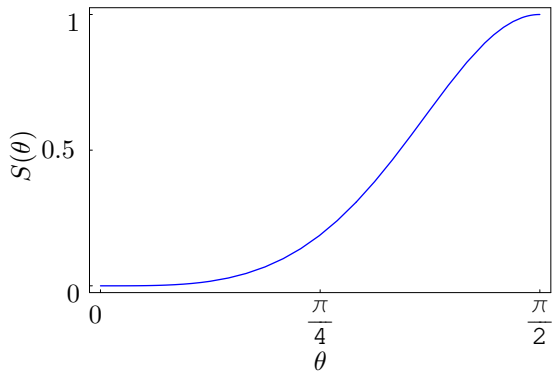}
\caption{(Color online) EoE $S(\theta)$ as a function of
$\theta$.}\label{figsse2}
\end{figure}

We plot in Figure \ref{figsse2} the EoE \cite{TK05}
$S(\theta)=-f_+(\theta)^2\log_2f_+(\theta)^2-f_-(\theta)^2\log_2f_-(\theta)^2$
of state (\ref{eqsse13}) as a function of $\theta$, for
$0<\theta\leq\pi/2$. The entanglement grows monotonically until
$\theta=\pi/2$, where it becomes maximal (1 e-bit).
\section{$\theta$-dependence of Bell's inequality violation}
In order to analyze the role the $\theta$ scattering angle plays in
the generation of these genuine quantum correlations, we consider
now the degree of violation of Bell's inequality as a function of
$\theta$. To this purpose, we define \cite{Bell64,CS78} the
observable
\begin{eqnarray}
E(\mathbf{\hat{a}},\mathbf{\hat{b}})&:=&\langle\Phi|(\mathbf{\sigma}^{(1)}\cdot
\mathbf{\hat{a}}\otimes\mathbf{\sigma}^{(2)}\cdot
\mathbf{\hat{b}})|\Phi\rangle\label{eqsse15}\\&=&-[\hat{a}_z\hat{b}_z+2f_p(\theta)f_p(\pi-\theta)(\hat{a}_x\hat{b}_x+\hat{a}_y\hat{b}_y)],\nonumber
\end{eqnarray}
where $|\Phi\rangle:=|\Phi_{\rm out}(\theta)\rangle$ is the
(normalized) state (\ref{j8}) and $\mathbf{\hat{a}}$,
$\mathbf{\hat{b}}$ are arbitrary unit vectors. In Eq.
(\ref{eqsse15}) we consider the amplitudes $f_p(\theta)$ and
$f_p(\pi-\theta)$ normalized for each $\theta$, in the form
$|f_p(\theta)|^2+|f_p(\pi-\theta)|^2=1$. We consider three coplanar
unit vectors, $\mathbf{\hat{a}}$, $\mathbf{\hat{b}}$ and
$\mathbf{\hat{c}}$.
$(\widehat{\mathbf{\hat{a}},\mathbf{\hat{b}}})=\pi/3$,
$(\widehat{\mathbf{\hat{a}},\mathbf{\hat{c}}})=2\pi/3$ and
$(\widehat{\mathbf{\hat{b}},\mathbf{\hat{c}}})=\pi/3$. We have
\begin{eqnarray}
&&|E(\mathbf{\hat{a}},\mathbf{\hat{b}})-E(\mathbf{\hat{a}},\mathbf{\hat{c}})|=1,\nonumber\\
&&F(\theta):=1+E(\mathbf{\hat{b}},\mathbf{\hat{c}})=\frac{5}{4}-\frac{3}{2}f_p(\theta)f_p(\pi-\theta).
\end{eqnarray}
The Bell's inequality, given by \cite{Bell64,CS78}
\begin{eqnarray}
|E(\mathbf{\hat{a}},\mathbf{\hat{b}})-E(\mathbf{\hat{a}},\mathbf{\hat{c}})|\leq
1+E(\mathbf{\hat{b}},\mathbf{\hat{c}}),
\end{eqnarray}
will then be
\begin{eqnarray}
F(\theta)\geq 1.
\end{eqnarray}

\begin{figure}
\includegraphics{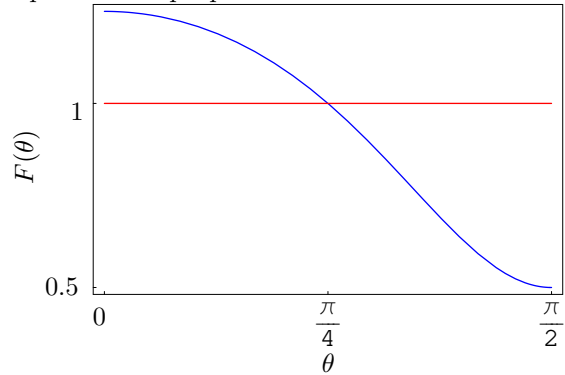}
\caption{(Color online) $F(\theta)$ as a function of $\theta$. The
classical-quantum border corresponds to $F(\theta_c)=1$, with
$\theta_c=\pi/4$.}\label{figsse3}
\end{figure}

For the particular case of Coulomb interaction at lowest order here
considered,
$2f_p(\theta)f_p(\pi-\theta)=2f_+(\theta)f_-(\theta)=(1-\cos^2\theta)/(1+\cos^2\theta)$
and thus the critical angle for which the inequality becomes
violated is $\theta_c=\pi/4$ for $F(\theta_c)=1$. For
$\theta_c<\theta\leq\pi/2$ the Bell's inequality does not hold.
 We show in Figure \ref{figsse3} the $\theta$ dependence
of $F(\theta)$ together with the classical-quantum border, $F=1$, at
$\theta_c=\pi/4$. Thus, for experiments with $\theta>\pi/4$ one
could be able in principle to discriminate between local realism and
quantum mechanics.
This is in contrast with recent analysis of Bell's inequalities
violation in elementary particle systems \cite{Go04,Bra05,Tor86},
where the emphasis was placed on flavor entanglement,
$K^0\bar{K}^0$, $B^0\bar{B}^0$, and the like. These analysis
presented \cite{Bra04} two kinds of drawbacks coming from the lack
of experimenter's free will and from the unitary evolution with
decaying states. These issues reduce the significance of the
experiments up to the point of preventing their use as tests of
quantum mechanics versus local realistic theories. The spin-spin
entanglement analyzed in this paper does not have this kind of
problems and could be used in principle for that purpose.
\section{Conclusions}
In summary, we analyzed the relation between entanglement and
antisymmetrization for identical particles, in the context of
spin-independent particle scattering. We showed that, in order to
create genuine spin-spin quantum correlations between two $s=\half$
fermions, spin-dependent interactions are not compulsory. The
identity of the particles along with an interaction between degrees
of freedom different from the spin, suffice for this purpose. The
entanglement generated this way is not a fictitious one due to
antisymmetrization, but a real one, and violates a certain Bell's
inequality for $\theta>\theta_c=\pi/4$.
\section*{ACKNOWLEDGMENTS}
This work was partially supported by the Spanish MEC project No.
FIS2005-05304. L.L. acknowledges support from the FPU grant No.
AP2003-0014.

\end{document}